\documentclass[reprint,aps]{revtex4-1}
\usepackage{graphicx,epstopdf,float,setspace,mathtools}
\pdfoutput=1
\begin{document}
\title{Discrete Density of States}
\thanks{NOTICE: This is the author’s version of a work that was accepted for publication in Physics Letters A. Changes resulting from the publishing process, such as peer review, editing, corrections, structural formatting, and other quality control mechanisms may not be reflected in this document. Changes may have been made to this work since it was submitted for publication. A definitive version will subsequently be published in Phys. Lett. A. DOI: 10.1016/j.physleta.2016.01.034}
\author{Alhun Aydin}
\author{Altug Sisman}
 \email{Corresponding author, sismanal@itu.edu.tr}
\affiliation{Nano Energy Research Group, Energy Institute, Istanbul Technical University, 34469, Istanbul, Turkey\\}
\date{\today}
\begin{abstract}
By considering the quantum-mechanically minimum allowable energy interval, we exactly count number of states (NOS) and introduce discrete density of states (DOS) concept for a particle in a box for various dimensions. Expressions for bounded and unbounded continua are analytically recovered from discrete ones. Even though substantial fluctuations prevail in discrete DOS, they're almost completely flatten out after summation or integration operation. It's seen that relative errors of analytical expressions of bounded/unbounded continua rapidly decrease for high NOS values (weak confinement or high energy conditions), while the proposed analytical expressions based on Weyl's conjecture always preserve their lower error characteristic.
\end{abstract}
\maketitle
\section{Introduction}
Density of states (DOS) is a useful concept that is extensively used in condensed matter and statistical physics. Although being a well-established and widely used concept, validity of conventional DOS is restricted by unbounded continuum approximation. Due to the rapid development of nanoscience and nanotechnology in recent years, detailed examination of DOS concept which is still commonly used in those areas became a necessity \cite{zhangbook,qbook,pathbook,nanotermo,ref7,ref10,ieee}. Moreover, advances in computational power of computers made it possible to exactly calculate the summations representing physical quantities, which is previously hard to do \cite{ieee,aydin}.

Essentially, state space is always discrete due to the finite size of domains and the wave character of particles. However, discreteness is usually neglected in case of the domain size is much larger than the de Broglie wavelength of particles, namely in macroscale. This leads to a continuous DOS (CDOS) function that is commonly used in literature. On the other hand, in nanoscale, at least one of the domain sizes is in the order of the de Broglie wavelength of particles and for such confined domains bounded continuum approximation represents the state space more properly. In a confined domain, bounded continuum approximation considers the non-zero value of ground states of momentum components while still neglecting their discrete nature. In this regard, Weyl's conjecture for the asymptotic behavior of eigenvalues uses bounded continuum approximation and offers a more precise enumeration of states; thus it gives a more accurate DOS function which may be called here Weyl DOS (WDOS).

Both CDOS and WDOS functions are based on continuum approximation, since they use infinitesimal energy interval assumption. However, quantum-mechanically minimum allowable energy interval is finite, and discrete nature of state space becomes significant when quantum confinement is strong. In this case, exact DOS function can only be defined by considering discrete energy eigenvalues. This treatment allows us to define discrete DOS (DDOS) function.

Long-standing unsolved Gauss' circle problem (or sphere in 3D case) that asks an analytical answer for how many integer lattice points inside of a circle with a given radius is profoundly related to the exact calculation of number of states in state space. Even though Gauss' circle and sphere problems are studied extensively for many years, still there are no exact analytical solutions in terms of elementary functions for both problems \cite{ref2,ref3,sumbook}. There are some studies related to these problems in literature for calculation of lattice sums \cite{sumbook,ref5,ref9,ref11,ref13}. Only a limited number of studies consider the evaluation of DOS functions for finite-size systems \cite{ref7,ref10,ieee,ref2,ref1,ref6,weylfr,weylfr2,ref12}. On the other hand, they use approximations and assumptions instead of considering the exact energy interval to define DOS function. Also, none of them give an exact and discrete DOS function for a particle in a box which is one of the most fundamental models used in statistical physics.

The aim of this Letter is, to introduce a discrete density of states function and to compare its results with conventional CDOS as well as Weyl's conjecture-based WDOS function that is proposed here. DDOS function is based on the exact enumeration of number of states (NOS) for quantum-mechanically allowable discrete energy levels, instead of using an infinitesimally small energy interval concept. In that sense, DDOS is the generalized form of DOS function which reduces to WDOS and CDOS functions in bounded and unbounded continuum limits respectively.

\section{Generalized forms of DDOS, WDOS and CDOS functions for a particle in a box}
As it is commonly preferred during the derivation of DOS in literature, we consider a non-interacting and non-relativistic massive particle confined in a $D$-dimensional rectangular domain. Dimensionless translational energy eigenvalues from the solution of Schr\"odinger equation for this kind of system are
\begin{equation}
\tilde{\varepsilon}=\frac{\varepsilon}{k_B T}=\frac{h^2}{8mk_BT}\sum_{n=1}^{D}{\left(\frac{i_n}{L_n}\right)^2}=\sum_{n=1}^{D}{(\alpha_n i_n)^2}
\end{equation}
where $D$ is the number of spatial dimensions, $k_B$ is Boltzmann's constant, $T$ is temperature, $h$ is Planck's constant, $m$ is single particle mass, $n$ denotes orthogonal directions, $L_n$ is length of the domain in $n$'th direction and $i_n$ is quantum state variable running from one to infinity. For convenience, we define here a confinement parameter $\alpha$ as $\alpha_n=h/\left(\sqrt{8mk_B T}L_n\right)$ to indicate the magnitude of confinement of the domain in direction $n$. It should be noted that, we use dimensionless energy throughout the derivations, $\tilde{\varepsilon}=\varepsilon/k_B T$, instead of energy itself for the simplicity of operations and the compactness of obtained expressions.

Let $f$ be a Lebesque-integrable function representing the physical quantity to be calculated. Summation of $f$ over all accessible quantum states gives the physical quantity for the system. Apart from some exceptional cases, exact results of sums cannot be given analytically but only numerically. On the other hand, as long as confinement parameters are much smaller than unity, sums can be replaced by integrals with a negligible error, and thus analytical results can be obtained. Multiple sums turn into multiple integrals and CDOS function allows to calculate these multiple integrals over quantum state variables by a single integral over energy states,
\begin{equation}
\int_0^{\infty}\mkern-14mu\cdots\int_0^{\infty}\mkern-8mu{f(\tilde{\varepsilon}_{i_1,\cdots i_D})di_1\cdots di_D}=\int_0^{\infty}\mkern-8mu{f(\tilde{\varepsilon})\textit{CDOS}(\tilde{\varepsilon})d\tilde{\varepsilon}}
\end{equation}
where $\textit{CDOS}(\tilde{\varepsilon})=d\Omega_D/d\tilde{\varepsilon}$, $d\Omega_D$ is the number of states having energy values between $\tilde{\varepsilon}$ and $\tilde{\varepsilon}+d\tilde{\varepsilon}$ in $D$-dimensional space and $d\tilde{\varepsilon}$ is the infinitesimal energy interval.

On the contrary, when the confinement parameters are close to or even exceed unity, deviations between the results of integrals and sums become important. In this case, multiple summations may need to be exactly calculated instead of their integral approximations and DDOS function allows to calculate multiple summations by a single summation as long as energy eigenvalues are explicitly known. In that case, usage of DDOS function is given as follows
\begin{equation}
\sum_{i_1=1}^{\infty}\cdots\sum_{i_D=1}^{\infty}{f(\tilde{\varepsilon}_{i_1,\cdots i_D})\Delta i_1\cdots \Delta i_D}=\sum_{\tilde{\varepsilon}=\tilde{\varepsilon}_0}^{\infty}f(\tilde{\varepsilon})\textit{DDOS}(\tilde{\varepsilon})\Delta\tilde{\varepsilon}
\end{equation}
where $\tilde{\varepsilon}_0=\alpha_1^2+\cdots+\alpha_D^2$ is ground state energy and $\Delta\tilde{\varepsilon}$ is the quantum-mechanically minimum allowable difference between successive energy levels, which is not a constant, unlike $d\tilde{\varepsilon}$. Unfortunately, it is not possible to obtain an analytical expression for $\Delta\tilde{\varepsilon}$ except for 1D case. Therefore, it is necessary to generate energy spectrum data by using Eq. (1) and apply ascending sorting process to this data, then calculate the exact energy intervals between successive energy levels numerically. Consequently, DDOS can be defined as,
\begin{equation}
\textit{DDOS}_{D}(\tilde{\varepsilon})=\frac{\Delta\Omega_D(\tilde{\varepsilon})}{\Delta\tilde{\varepsilon}}=\frac{\Omega_{D}(\tilde{\varepsilon}+\Delta\tilde{\varepsilon})-\Omega_{D}(\tilde{\varepsilon})}{\Delta\tilde{\varepsilon}}
\end{equation}
where $\Omega_D$ is discrete number of states (DNOS) given by,
\begin{equation}
\Omega_D(\tilde{\varepsilon})=\textit{DNOS}_{D}(\tilde{\varepsilon})=\sum_{i_1^{\prime}=1}^{\infty}\cdots\sum_{i_D^{\prime}=1}^{\infty}\Theta\left[\tilde{\varepsilon}-\sum_{n=1}^{D}{(\alpha_n i_n^{\prime})^2}\right]
\end{equation}
where $\Theta$ is left-continuous Heaviside step function, $\Theta(0)=0$. It is clear that the difference of number of states for two successive energy levels ($\tilde{\varepsilon}$ and $\tilde{\varepsilon}+\Delta\tilde{\varepsilon}$) equal to the degeneracy of the energy level $\tilde{\varepsilon}$ since there are no states located in between successive energy levels. Note that, we consider spinless particles for brevity since spin degree of freedom is just a multiplication constant.

DDOS function predicts some exceptional results than those of CDOS function and it gives deeper physical insights which can be used in physical interpretations of non-trivial behaviors appeared in confined structures. While DDOS function gives an exact description for DOS function, it requires to know the shape of the domain and calculate the energy eigenvalues explicitly. In order to obtain an approximate DOS function for an arbitrary-shaped domain, the best approximation is to use Weyl's conjecture derived under bounded continuum approximation by neglecting discreteness. Weyl's conjecture that gives the asymptotic behavior of the number of eigenvalues less than $k$ for Helmholtz wave equation (which is the stationary form of Schro\"dinger equation for a particle in a box) in a $D$-dimensional finite-size domain is commonly written as \cite{pathbook},
\begin{equation}
\begin{split}
\Omega(k)= &\frac{Vk^3}{6\pi^2}\Theta(D-2)+(-1)^D\frac{Sk^2}{4^{D-2}4\pi}\Theta(D-1) \\
& +(-1)^{D-1}\frac{Pk}{4^{D-1}\pi}\Theta(D)+(-1)^{D-2}\frac{N_E}{4^D}
\end{split}
\end{equation}
where $k$ is wavenumber, $V$, $S$, $P$ and $N_E$ are volume, surface, periphery and number of edges of the domain respectively. By considering parabolic dispersion relation between $\varepsilon$ and $k$, we may obtain WNOS and WDOS functions respectively from Eq. (6) as,
\begin{equation}
\begin{split}
\textit{WNOS}_{D}(\tilde{\varepsilon})=
& \sum_{\tilde{\varepsilon}_s}\Biggl[\frac{4}{3\sqrt{\pi}}\frac{V}{\lambda_{th}^3}\tilde{\varepsilon}^{3/2}\Theta(D-2) \\
& +\frac{(-1)^D}{4^{D-2}}\frac{S}{\lambda_{th}^2}\tilde{\varepsilon}\Theta(D-1) \\
& +\frac{(-1)^{D-1}}{4^{D-1}}\frac{2}{\sqrt{\pi}}\frac{P}{\lambda_{th}}\sqrt{\tilde{\varepsilon}}\Theta(D)+\frac{(-1)^{D-2}}{4^{D}}N_E\Biggr]
\end{split}
\end{equation}
\begin{equation}
\begin{split}
\textit{WDOS}_{D}(\tilde{\varepsilon})=
& \sum_{\tilde{\varepsilon}_s}\Biggl[\frac{2}{\sqrt{\pi}}\frac{V}{\lambda_{th}^3}\sqrt{\tilde{\varepsilon}}\Theta(D-2)+\frac{(-1)^D}{4^{D-2}}\frac{S}{\lambda_{th}^2}\Theta(D-1) \\
& +\frac{(-1)^{D-1}}{4^{D-1}}\frac{1}{\sqrt{\pi}}\frac{P}{\lambda_{th}}\frac{1}{\sqrt{\tilde{\varepsilon}}}\Theta(D)\Biggr]
\end{split}
\end{equation}
where $\lambda_{th}=h/\sqrt{2\pi m k_B T}$ is thermal de Broglie wavelength, Heaviside step functions are left-continuous and $\tilde{\varepsilon}_s$ represents energy eigenvalues of subbands. Subbands are associated with quantized modes for confined directions of a domain. Hence, the number of confined directions denote the number of subband summations. \textit{e.g.}, if the first direction is confined while the other two are free (quasi-2D), then $\tilde{\varepsilon}_s=(\alpha_1 i_1^{\prime})^2$ and there is one summation over $i_1^{\prime}$; but if the first and second directions are confined and the other one is free (quasi-1D), then $\tilde{\varepsilon}_s=(\alpha_1 i_1^{\prime})^2+(\alpha_2 i_2^{\prime})^2$ and there is a double summation over $i_1^{\prime}$ and $i_2^{\prime}$.

CNOS and CDOS functions can be recovered from Eqs. (7) and (8) by neglecting higher order terms and considering rectangular domain geometry as follows,
\begin{equation}
\textit{CNOS}_{D}(\tilde{\varepsilon})=\sum_{\tilde{\varepsilon}_s}\frac{\pi^{D/2}\Theta\left(\tilde{\varepsilon}-\tilde{\varepsilon}_s\right)}{2^D\Gamma[(D+2)/2]}\frac{\left(\tilde{\varepsilon}-\tilde{\varepsilon}_s\right)^{D/2}}{\alpha_1\cdots\alpha_D}
\end{equation}
\begin{equation}
\textit{CDOS}_{D}(\tilde{\varepsilon})=\sum_{\tilde{\varepsilon}_s}\frac{\pi^{D/2}\Theta\left(\tilde{\varepsilon}-\tilde{\varepsilon}_s\right)}{2^D \Gamma[D/2]}\frac{\left(\tilde{\varepsilon}-\tilde{\varepsilon}_s\right)^{(D-2)/2}}{\alpha_1\cdots\alpha_D}
\end{equation}
respectively. In the following section, we examine DDOS function for structures with various confined dimensions.

\section{Derivations of WDOS and CDOS functions from DDOS in various spatial dimensions}
DDOS functions are need to be numerically calculated by using Eqs. (4) and (5). On the other hand, WDOS (and CDOS) functions can analytically be obtained by considering bounded (and unbounded) continuum approximations and Poisson summation formula (PSF) \cite{mathbook}.
\subsection{3D DOS}
In order to obtain DOS function, it is necessary to calculate NOS function first. Enumeration of number of discrete states up to an energy level ($\tilde{\varepsilon}$) can exactly be done by using Eq. (5) for 3D case,
\begin{equation}
\textit{DNOS}_3(\tilde{\varepsilon})=\sum_{i_1^{\prime}=1}^{\infty}\sum_{i_2^{\prime}=1}^{\infty}\sum_{i_3^{\prime}=1}^{\infty}\Theta(i_1^{\prime},i_2^{\prime},i_3^{\prime})
\end{equation}
where $\Theta(i_1^{\prime},i_2^{\prime},i_3^{\prime})=\Theta[\tilde{\varepsilon}-\left(\alpha_1 i_1^{\prime}\right)^2-\left(\alpha_2 i_2^{\prime}\right)^2-\left(\alpha_3 i_3^{\prime}\right)^2]$.

To yield analytical expressions corresponding to bounded continuum, $\{\alpha_1,\alpha_2,\alpha_3\}<<1$, we may apply the first two terms of PSF for the sums in Eq. (11) as follows,
\begin{equation}
\begin{split}
& \textit{DNOS}_3\approx\int\limits_0^{\infty}\int\limits_0^{\infty}\int\limits_0^{\infty}\Theta_3(i_1^{\prime},i_2^{\prime},i_3^{\prime})di_1^{\prime}di_2^{\prime}di_3^{\prime} \\
& -\frac{1}{2}\Biggl[\int\limits_0^{\infty}\int\limits_0^{\infty}\Theta_3(i_1^{\prime},i_2^{\prime},0)di_1^{\prime}di_2^{\prime}+\int\limits_0^{\infty}\int\limits_0^{\infty}\Theta_3(i_1^{\prime},0,i_3^{\prime})di_1^{\prime}di_3^{\prime} \\
& +\int\limits_0^{\infty}\int\limits_0^{\infty}\Theta_3(0,i_2^{\prime},i_3^{\prime})di_2^{\prime}di_3^{\prime}\Biggr]+\frac{1}{4}\Biggl[\int\limits_0^{\infty}\Theta_3(i_1^{\prime},0,0)di_1^{\prime} \\
& +\int\limits_0^{\infty}\Theta_3(0,i_1^{\prime},0)di_2^{\prime}+\int\limits_0^{\infty}\Theta_3(0,0,i_3^{\prime})di_3^{\prime}\Biggr]-\frac{1}{8} \\
& =\frac{\pi\tilde{\varepsilon}^{3/2}}{6\alpha_1\alpha_2\alpha_3}-\frac{\pi\tilde{\varepsilon}}{8}\left(\frac{1}{\alpha_1\alpha_2}+\frac{1}{\alpha_1\alpha_3}+\frac{1}{\alpha_2\alpha_3}\right) \\
& +\frac{\sqrt{\tilde{\varepsilon}}}{4}\left(\frac{1}{\alpha_1}+\frac{1}{\alpha_2}+\frac{1}{\alpha_3}\right)-\frac{1}{8}=\textit{WNOS}_3
\end{split}
\end{equation}
Here we obtained an analytical expression for 3D NOS that contains surface (terms along with the first bracket), periphery (terms along with the second bracket) and edge (the last term) corrections of Weyl's conjecture. Therefore, this expression may be called WNOS. When confinement parameters are much less than unity, finite-energy difference, $\Delta\tilde{\varepsilon}$, can fairly be approximated by infinitesimal energy difference, $d\tilde{\varepsilon}$. In that case, operation represented by Eq. (4) becomes an ordinary differentiation and WDOS can directly be obtained from Eq. (12) by differentiation,
\begin{equation}
\begin{split}
\textit{WDOS}_3=
& \frac{\pi\sqrt{\tilde{\varepsilon}}}{4\alpha_1\alpha_2\alpha_3}-\frac{\pi}{8}\left(\frac{1}{\alpha_1\alpha_2}+\frac{1}{\alpha_1\alpha_3}+\frac{1}{\alpha_2\alpha_3}\right) \\
& +\frac{1}{8\sqrt{\tilde{\varepsilon}}}\left(\frac{1}{\alpha_1}+\frac{1}{\alpha_2}+\frac{1}{\alpha_3}\right)
\end{split}
\end{equation}
When $\{\alpha_1,\alpha_2,\alpha_3\}\rightarrow 0$, the contribution of the first term becomes dominant and Eq. (13) reduces to
\begin{equation}
\begin{split}
\textit{CDOS}_3=\frac{\pi\sqrt{\tilde{\varepsilon}}}{4\alpha_1\alpha_2\alpha_3}
\end{split}
\end{equation}
which is the well-known CDOS expression for 3D case.

\subsection{2D DOS}
Discrete number of states in 2D case is represented by,
\begin{equation}
\textit{DNOS}_2=\sum_{i_1^{\prime}=1}^{\infty}\sum_{i_2^{\prime}=1}^{\infty}\Theta\left[\tilde{\varepsilon}-\left(\alpha_1 i_1^{\prime}\right)^2-\left(\alpha_2 i_2^{\prime}\right)^2\right]
\end{equation}

We may obtain an analytical NOS formula for bounded continuum, $\{\alpha_1,\alpha_2\}<<1$, containing Weyl's corrections, by following similar procedures in 3D case for summations in Eq. (15),
\begin{equation}
\begin{split}
& \textit{DNOS}_2\approx\int\limits_0^{\infty}\int\limits_0^{\infty}\Theta\left[\tilde{\varepsilon}-\left(\alpha_1 i_1^{\prime}\right)^2-\left(\alpha_2 i_2^{\prime}\right)^2\right]di_1^{\prime}di_2^{\prime} \\
& -\frac{1}{2}\int\limits_0^{\infty}\Theta\left[\tilde{\varepsilon}-\left(\alpha_1 i_1^{\prime}\right)^2\right]di_1^{\prime}-\frac{1}{2}\int\limits_0^{\infty}\Theta\left[\tilde{\varepsilon}-\left(\alpha_2 i_2^{\prime}\right)^2\right]di_2^{\prime} \\
& +\frac{1}{4}\Theta\left[\tilde{\varepsilon}\right]=\frac{\pi\tilde{\varepsilon}}{4\alpha_1\alpha_2}-\frac{\sqrt{\tilde{\varepsilon}}}{2\alpha_1}-\frac{\sqrt{\tilde{\varepsilon}}}{2\alpha_2}+\frac{1}{4}=\textit{WNOS}_2
\end{split}
\end{equation}
which is the WNOS function for 2D. Taking the derivative of the result in Eq. (16) with respect to energy gives the WDOS,
\begin{equation}
\begin{split}
\textit{WDOS}_2=\frac{\pi}{4\alpha_1\alpha_2}-\frac{1}{4\sqrt{\tilde{\varepsilon}}}\left(\frac{1}{\alpha_1}+\frac{1}{\alpha_2}\right)
\end{split}
\end{equation}
The second terms are inversely proportional to energy and represent the peripheral corrections in Weyl's conjecture. Although energy independence is a well-known property of 2D DOS function in unbounded continuum case, it becomes energy-dependent in bounded continuum case even if discrete corrections are still negligible. In infinite domains, correction terms in Eq. (17) also looses their importance and we recover the CDOS function which is independent of energy,
\begin{equation}
\begin{split}
& \textit{CDOS}_2=\frac{\pi}{4\alpha_1\alpha_2}
\end{split}
\end{equation}
that is actually the term obtained from double integral in Eq. (16).

\subsection{1D DOS}
Similar to 3D and 2D cases, DNOS for 1D is
\begin{equation}
\textit{DNOS}_1=\sum_{i_1^{\prime}=1}^{\infty}\Theta\left[\tilde{\varepsilon}-\left(\alpha_1 i_1^{\prime}\right)^2\right]
\end{equation}
Unlike in 3D and 2D cases, it is possible to get an analytical expression for DDOS. It is obvious that there is only one state in each energy level, $\Delta\Omega_1=\Omega_{1}(\tilde{\varepsilon}+\Delta\tilde{\varepsilon})-\Omega_{1}(\tilde{\varepsilon})=\Delta i=1$ for 1D case. Since there is no degeneracy of energy states in 1D case, exact energy interval can be analytically expressed as $\Delta\tilde{\varepsilon}=2\alpha_1^2 i_1+\alpha_1^2$. Consequently, DDOS can analytically be given by
\begin{equation}
\textit{DDOS}_1=\frac{\Delta\Omega_1}{\Delta\tilde{\varepsilon}}=\frac{1}{2\alpha_1 \sqrt{\tilde{\varepsilon}}+\tilde{\varepsilon}_0}
\end{equation}
Approximation of DNOS by PSF for bounded 1D continuum gives,
\begin{equation}
\begin{split}
\textit{DNOS}_1 &
\approx\int_0^{\infty}\Theta\left[\tilde{\varepsilon}-\left(\alpha_1 i_1^{\prime}\right)^2\right]di_1^{\prime}-\frac{1}{2}\Theta\left[\tilde{\varepsilon}\right] \\
& =\frac{\sqrt{\tilde{\varepsilon}}}{\alpha_1}-\frac{1}{2}=\textit{WNOS}_1
\end{split}
\end{equation}
Second term represents the exclusion of the false contribution of non-existing zeroth state in the integral.

Derivative of Eq. (21) gives
\begin{equation}
\textit{WDOS}_1=\frac{1}{2\alpha_1 \sqrt{\tilde{\varepsilon}}}=\textit{CDOS}_1
\end{equation}
which means WDOS and CDOS are equal in 1D case, although NOS functions are different. It's seen from Eqs. (20) and (22) that, DDOS/CDOS reaches to its minimum value $(2/3)$ at ground state energy ($\tilde{\varepsilon}=\tilde{\varepsilon}_0=\alpha_1^2$) in where the error of CDOS is maximum. In fact, for non-zero energy eigenvalues in a free domain where confinement parameters go to zero, $\tilde{\varepsilon}>>\tilde{\varepsilon}_0$ condition can be fulfilled since ground state energy goes to zero ($\tilde{\varepsilon}_0\rightarrow 0$) and DDOS is directly reduced to CDOS.

\subsection{0D DOS}
Although it may be seen as trivial, it is traditional to discuss the zero-dimensional case also. In 0D case, the only available state is the ground state, $\tilde{\varepsilon}_0$. Therefore, from Eq. (19), DNOS can be written as
\begin{equation}
\textit{DNOS}_0=\Theta(\tilde{\varepsilon}-\tilde{\varepsilon}_0)
\end{equation}
Since there is no summation, DNOS is inherently equal to WNOS and also to CNOS, $\textit{DNOS}_0=\textit{WNOS}_0=\textit{CNOS}_0$. Hence, DDOS can directly be obtained by differentiating Eq. (23) with respect to $\tilde{\varepsilon}$,
\begin{equation}
\textit{DDOS}_0=\frac{d \textit{DNOS}_0(\tilde{\varepsilon})}{d\tilde{\varepsilon}}=\delta(\tilde{\varepsilon}-\tilde{\varepsilon}_0)=\textit{WDOS}_0=\textit{CDOS}_0.
\end{equation}

\section{Results and Discussion}

\begin{figure}[t]
\centering
\includegraphics[width=0.48\textwidth]{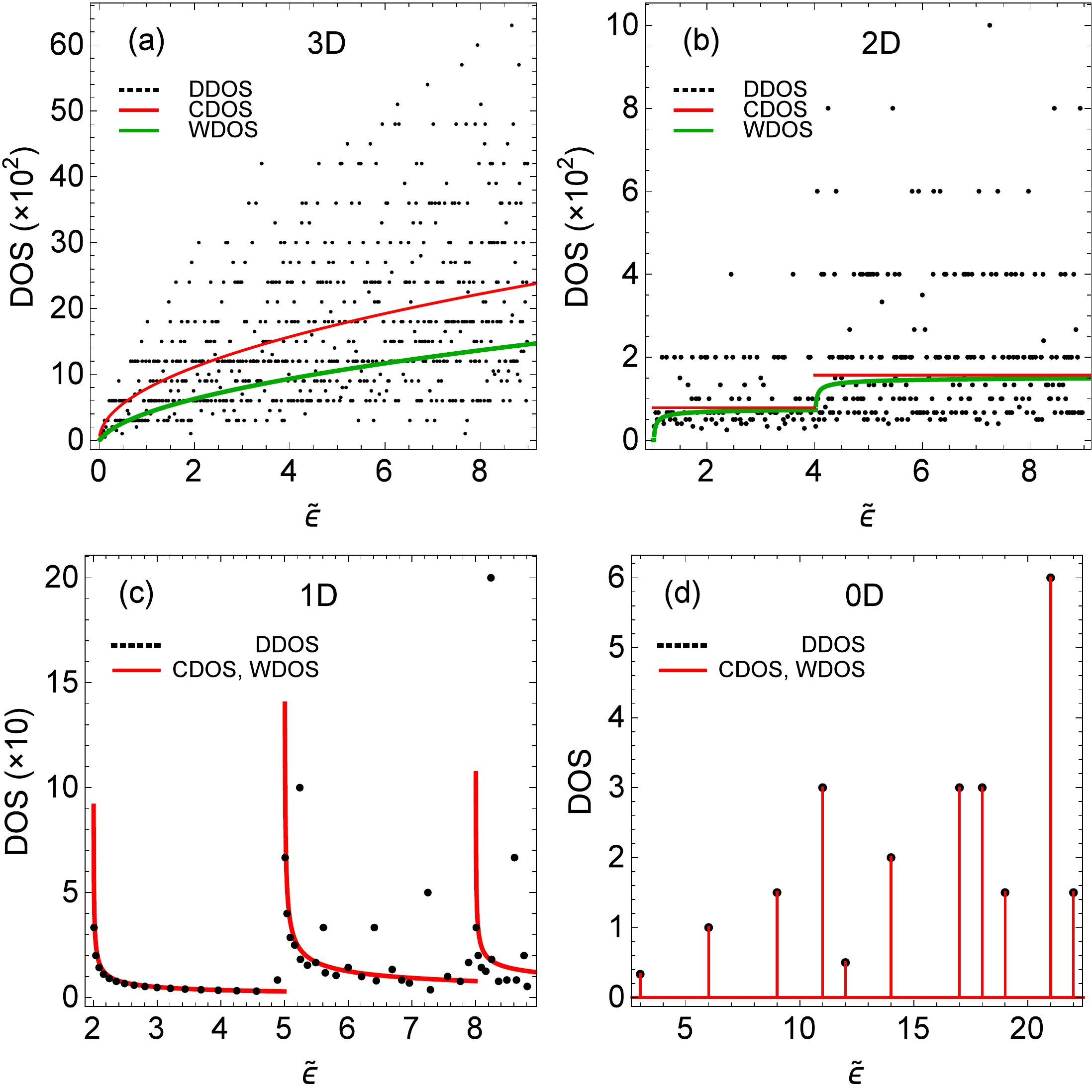}
\caption{DDOS, CDOS and WDOS functions varying with energy. (a) 3D box with $\alpha_1=\alpha_2=\alpha_3=0.1$, (b) Quasi-2D quantum well with $\alpha_1=\alpha_2=0.1, \alpha_3=1$, (c) Quasi-1D quantum wire with $\alpha_1=0.1, \alpha_2=\alpha_3=1$ and (d) Quasi-0D quantum dot with $\alpha_1=\alpha_2=\alpha_3=1$.}
\label{fig:pic1}
\end{figure}

Comparisons of discrete, Weyl and continuous DOS functions are shown in Fig. (1) for various dimensions. It is seen that DDOS has an extremely fluctuating behavior although it roughly preserves the usual functional dependency on energy. In spite of the enormous fluctuations in DDOS, conventional CDOS is commonly used in calculations even at nanoscale. It has not been clearly examined or explained in literature that, how quantities calculated from CDOS in a certain extent correctly predicts the experimental results while CDOS and DDOS give radically different results. One may reasonably ask why huge fluctuations of DDOS does not affect the measured quantities. In fact, in almost all thermodynamic or transport quantities, DOS functions are either alone or multiplied by a smoothly varying functions in the kernels of integrals or summations. Accumulative nature of summation or integration process regulates the dispersive behavior, just like in NOS functions, Fig. (2). In other words, after summation or integration process, positive and negative deviations from WDOS (or CDOS) function substantially cancel each other. That's why fluctuations coming from the discrete nature of DDOS are almost completely flatten in DNOS functions in Fig. (2).

The reason of the difference between WNOS and CNOS is the false contributions from surface, line and edge modes in CNOS, which is corrected in WNOS. Therefore, instead of CNOS, another analytical function, WNOS, need to be used even in weakly confined structures. On the other hand, when DOS function is considered, interestingly least square errors of CDOS functions in Figs. (1a) and (1b) are less than those of WDOS. However, these errors have no importance since DOS functions are always used under summation or integration operators and after this accumulative operations WDOS gives always better results than CDOS, as seen in NOS functions in Fig. (2). Comparisons of successes of WNOS and CNOS functions in terms of relative errors are also given as sub-figures in Fig. (2). The errors of WNOS functions are considerably lower than those of CNOS functions. As the energy increases, errors are decreasing drastically.

\begin{figure}[t]
\centering
\includegraphics[width=0.48\textwidth]{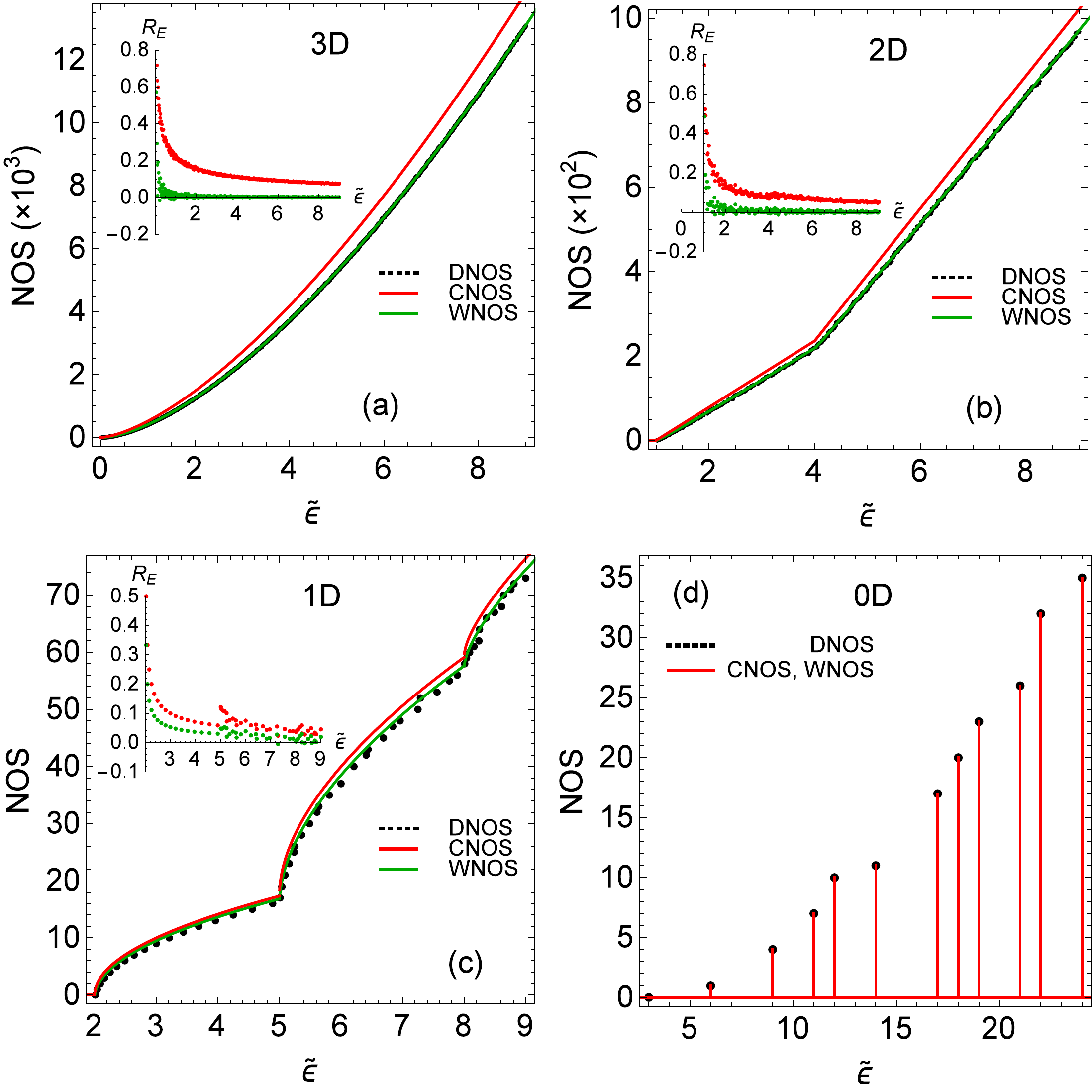}
\caption{DNOS, CNOS and WNOS functions and its relative errors varying with energy. (a) 3D box with $\alpha_1=\alpha_2=\alpha_3=0.1$, (b) Quasi-2D quantum well with $\alpha_1=\alpha_2=0.1, \alpha_3=1$, (c) Quasi-1D quantum wire with $\alpha_1=0.1, \alpha_2=\alpha_3=1$ and (d) Quasi-0D quantum dot with $\alpha_1=\alpha_2=\alpha_3=1$.}
\label{fig:pic1}
\end{figure}

In Fig. (3a), variations of relative errors of 3D CNOS and WNOS functions with energy for two different values of confinement parameters are shown in detail. It is clearly seen that the error of WNOS is much smaller than CNOS and errors of both functions decreases with increasing energy or decreasing confinement. Relative errors of CNOS and WNOS vs exact number of states (DNOS) is examined in Fig. (3b). Relative errors decrease rapidly with increasing NOS value while WNOS is preserving its high accuracy, see sub-figure of Fig. (3b). This situation does not depend on the values of confinement parameters.

States are 0D points in both momentum and quantum state spaces. During the derivations of CDOS and WDOS, the points in momentum state space are considered as if they are finite-size cells. In order to use this approximation, the highest energy level considered has to be far from the ground state. Otherwise, number of edges over periphery $(N_E/P)$, periphery over area $(P/A)$ and area over volume $(A/V)$ ratios will not be sufficiently small and because of that conversion of state points into finite-size cells can cause considerably large amount of errors in DOS and NOS functions. The first condition to use the continuum approximation is then $\tilde{\varepsilon}>>\tilde{\varepsilon}_0$ regardless of the size of the cell (values of confinement parameter $\alpha$'s). The second condition is $\alpha_n\rightarrow 0$ which indicates that the size of the cell is infinitesimally small, so that in the absence of confinement, cells turn into points indeed. However, the first condition still need to be satisfied, even if the second condition is already fulfilled. Otherwise, for the states near to ground state DDOS substantially deviates from CDOS even if confinement parameters go to zero. Thus, the first condition is more essential than the second one. In fact, these conditions can be represented by a unique condition of $\alpha_1\cdots\alpha_D/\tilde{\varepsilon}^{(D/2)}\rightarrow 0$, in which the expression is equal to reciprocal of CNOS except some constants around unity. Then, both conditions are intrinsically satisfied when $NOS\rightarrow\infty$, as it is clearly seen in Fig. (3b).

\begin{figure}[h]
\centering
\includegraphics[width=0.48\textwidth]{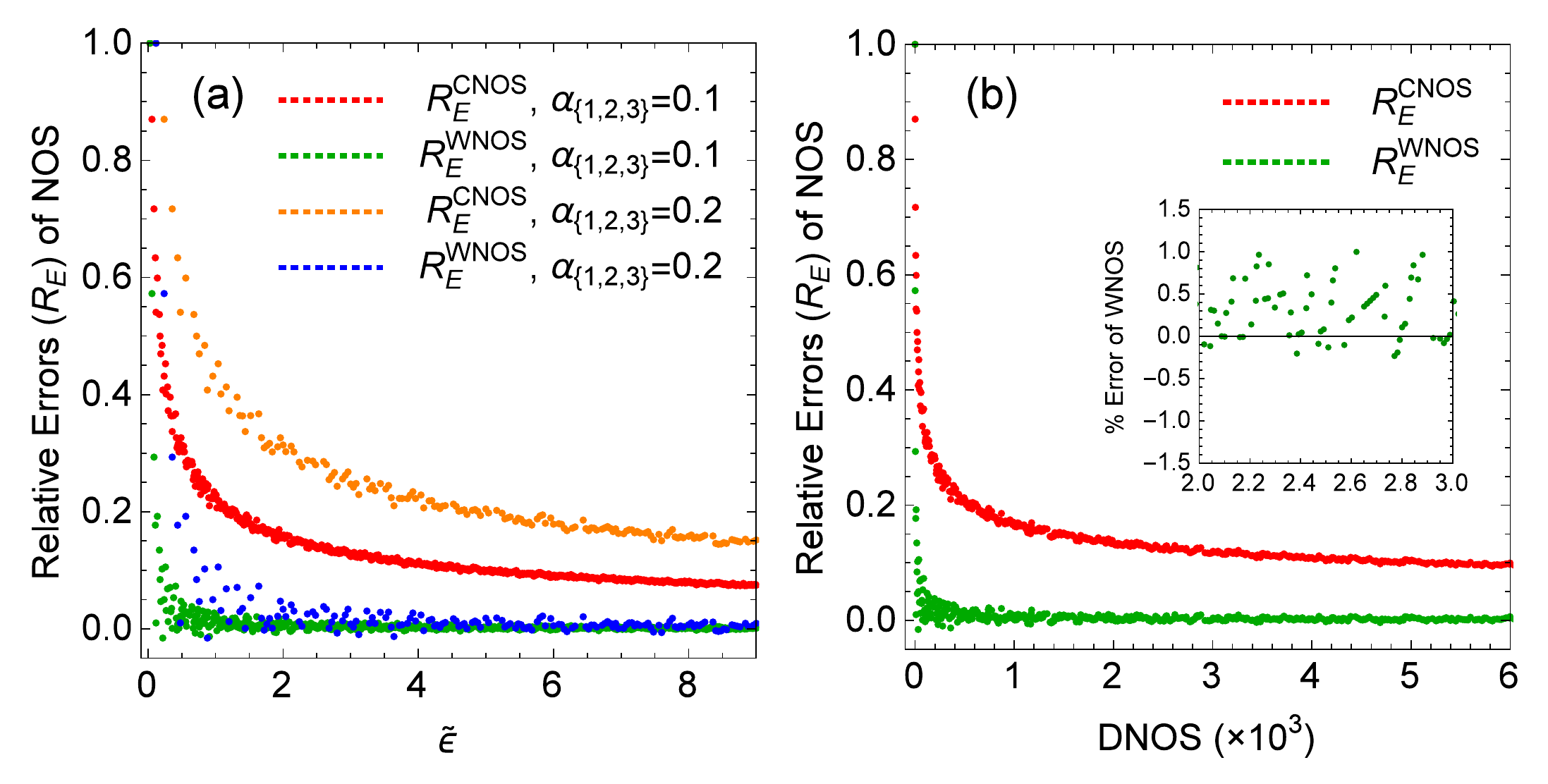}
\caption{Relative errors of WNOS and CNOS functions. (a) Comparisons for two different confinement values with changing energy, (b) Relative errors changing with exact number of states independent from the magnitudes of confinements. Sub-figure in (b) focuses on percentage errors of WNOS function for a particular DNOS range.}
\label{fig:pic1}
\end{figure}

\subsection{Dependence of dimensional behaviors on energy and aspect ratios}
Dimensionality behavior or order of energy dependence of DOS function can be changed by modifying aspect ratios (ratios of sizes in particular directions) of domains. If the sizes are the same at least in two directions, dimensional behavior of DOS function turns directly into the behavior of 3D, as the energy increased. Conversely, for completely anisometric domains with all sizes are strictly different from each other, energy increment leads to a gradual transition from lower to higher dimensions.

As is clearly seen from the analytical expressions in section 3., DOS (and NOS) has strictly different dependence on energy for various dimensions. We choose aspect ratios of confinement domains in Figs. (1) and (2) in a way that behaviors of functions at different dimensions become apparent. Any change in the magnitudes of confinement, while keeping aspect ratios the same, preserves the energy dependence patterns of DOS and NOS functions. For example, there is no difference in the fluctuation patterns of DDOS between $\alpha_{1,2,3}=0.01$ and $\alpha_{1,2,3}=1$, as long as we consider higher energy ranges, although they represent 3D and quasi-0D (for low energies) domains respectively. The only difference in between those cases is the energy scale. Change in the order of energy dependence of DOS function as a result of an energy increment is related with the number of isometrically confined directions. For example, a domain having quasi-1D behavior $(\alpha_1=0.01, \alpha_2=\alpha_3=1)$ for low energy values can start to represent 3D behavior for high enough energy values. For isometric domains (no preferred direction), change of quantum state variables in any direction requires the same energy change. Therefore, a quasi-0D domain starts to behave like 3D when the energy is increased enough.

\subsection{Conclusion}
In this Letter, by considering quantum-mechanically minimum allowable energy difference, we exactly counted the number of states and introduced DDOS concept. This treatment here can be seen as general for non-interacting and non-relativistic massive particles.

When the large portion of particles in a system occupy low energy levels near to ground state, ground state energy become dominant and deviations from continuum approximation become appreciable. Hence, at low temperatures or strong confinement conditions, in which occupation probabilities of low energy levels near to ground state are higher, DDOS may need to be used for more accurate calculations of physical properties. On the other hand, DDOS needs considerable computational calculations, although the amount of calculation becomes reasonable for nanoscale systems with energy near to ground state. A simpler way with an acceptable error (much less than conventional) is to use the proposed WDOS function to calculate physical quantities (like thermodynamic or transport quantities) of confined structures especially.

DDOS concept leads to a deeper understanding of the physical behaviors of matter and it can also serve as a helpful tool to understand especially the nanoscale behaviors of matter more accurately.

\bibliography{dosref}
\bibliographystyle{unsrt}
\end{document}